\documentclass[12pt]{article}
\usepackage{amssymb}
\usepackage{epsfig,caption,graphicx,eepic,epic}
\usepackage{pstricks}
\usepackage{amsmath}
\usepackage{multido}

\begin{document}
\def\be{\begin{equation}}
\def\ee{\end{equation}}
\def\ba{\begin{eqnarray}} 
\def\ea{\end{eqnarray}}
\def\nn{\nonumber}

\newcommand{\bbf}{\mathbf}
\newcommand{\rrm}{\mathrm}

\title{An application of renormalization in Hilbert space at phase transition points in strongly 
interacting systems\\} 

\author{Tarek Khalil
\footnote{E-mail address: khalil@lpt1.u-strasbg.fr}\\ 
and\\
Jean Richert
\footnote{E-mail address: richert@lpt1.u-strasbg.fr}\\ 
Laboratoire de Physique Th\'eorique, UMR 7085 CNRS/ULP,\\
Universit\'e Louis Pasteur, 67084 Strasbourg Cedex,\\ 
France} 
 
\date{\today}
\maketitle 
\begin{abstract}
We introduce an algorithm aimed to reduce the dimensions of Hilbert space. It is used here in 
order to study the behaviour of low energy states of strongly interacting quantum many-body 
systems at first order transitions and avoided crossings. The method is tested on different 
frustrated quantum spin ladders with two legs. The role and importance of symmetries are investigated 
by using different bases of states.
\end{abstract}


\section{Introduction.}

The investigation of the spectral properties of strongly interacting microscopic many-body 
quantum systems necessitates the use of non-perturbative approaches. Most of these rest
on the renormalization group concept introduced by Wilson ~\cite{wil} and universally used 
since in all fields of quantum physics. This is the case for the study of lattice quantum 
systems for which the Real Space Renormalization Group (RSRG)~\cite{mal,cap,whi} and the 
Density Matrix Renormalization Group (DMRG) ~\cite{malve,whi2,henk} have been introduced 
in order to reduce the dimensions of the systems. More generally the particular property of
the renormalization methods is the fact that they can reveal the existence and location of phase 
transitions which are characterized as fixed points, i.e. the coupling strengths which define 
the Hamiltonian or the action of the systems and evolve during the renormalization process 
(running coupling constants) stay constant at  these specific points, both in the case of 
discontinuous and continuous transitions~\cite{henk,sach1}.
 
The reduction process has mainly been applied in $d$-dimensional real and momentum 
space. In the present approach we introduce a reduction algorithm which operates in $0d$ 
Hilbert space. The spectral properties of quantum systems are obtained through the 
diagonalization of a many-body Hamiltonian acting in a complete, in general infinite or at
least very large set of basis states although the information of interest is restricted to the 
knowledge of a few low-energy states.

In order to avoid the diagonalization of very large matrices we recently proposed a 
non-perturbative approach which tackles this question ~\cite{khri}. We implemented it
in the study of strongly interacting systems like frustrated two-leg ladders~\cite{khri2}.
The approach consists of a step by step reduction of the size of Hilbert space by means of 
a projection technique. It induces a renormalization process in the spirit of former work
~\cite{gla,mue,bek}. Its advantage over other reduction procedures lies in the fact that 
it is applicable to all types of microscopic quantum systems since it works in Hilbert space
like in the procedure developed in ref.~\cite{bae}.

In the present work we concentrate on the use of the algorithm in order to characterise and 
study interacting systems in the vicinity of discontinuous (first order) phase transitions and 
avoided crossings which are signalized by the existence of fixed points at which coupling constants
stay constant during the reduction process. We investigate the ability of the algorithm to 
perform the reduction of the Hilbert space dimensions at the location of these fixed points
and to signalize the existence of phase transition points.

The outline of the paper is the following. In section $2$ we recall briefly the main lines of 
the space reduction algorithm which has been developed elsewhere ~\cite{khri} and show the implication 
of the existence of fixed points on the coupling strengths which enter the Hamiltonian of the system. 
In section $3$ we test the algorithm in the neighbourhood of fixed points which correspond to 
discontinuous transitions and avoided crossings in frustrated spin ladders. Conclusions are drawn 
in section $4$.  

\section{Formal algorithm and determination of fixed points}

\subsection{Space reduction procedure.}
  
\subsubsection{General concept}

We consider a system described by a Hamiltonian\\ 
$H^{(N)}\left ( g_1^{(N)},g_2^{(N)},...,g_p^{(N)}\right)$ which depends on $p$ coupling strengths\\
$\left\{ g_1^{(N)},g_2^{(N)},\cdots, g_p^{(N)}\mapsto g^{(N)}\right\}$ and acts in a Hilbert 
space ${\cal H}^{(N)}$  of dimension $N$.  $H^{(N)}$ has $N$ eigenvalues $\left\{\lambda_i(g^{(N)}) , 
\, i=1,\cdots, .N\right\}$  and eigenvectors $\left\{|\Psi_i^{(N)}(g^{(N)})\rangle,\, i=1,
\cdots,N\right\}$. 

If the relevant properties of the system are essentially located in a subspace ${\cal H}^{(M)} $  
of ${\cal H}^{(N)} $ ($M <N$) it makes sense to try to define a new effective Hamiltonian 
$H^{(M)}(g^{(M)})$ whose eigenvalues reproduce the selected subspace and verifies

\be\nonumber
H^{(M)}(g^{(M)}) |\Psi_i^{(M)}(g^{(M)})\rangle = \lambda_i(g^{(M)}) 
|\Psi_i^{(M)}(g^{(M)})\rangle  
\ee
with the constraints
\be
\lambda_i(g^{(M)}) = \lambda_i(g^{(N)})
\label{eq0} \ 
\ee
for $i = 1,...,M$. If this can be realized Eq.~(\ref{eq0}) implies a relation between the 
coupling constants in the original and reduced space
\be\nonumber
g_k ^{(M)} = f_k(g_1^{(N)},g_2^{(N)},...,g_p^{(N)})
\ee
with $k = 1,...,p$. We show below how this effective Hamiltonian $H^{(M)}(g^{(M)})$ can be 
constructed.

\subsubsection{Reduction algorithm and renormalization of the coupling strengths.}

Consider a system described by a Hamiltonian depending on a unique coupling strength 
$g$ which can be written as a sum of two terms 
\be
H = H_0 + H_1(g) 
\label{eq1} \ 
\ee

The Hilbert space  ${\cal H}^{(N)}$ of dimension $N$ is spanned by a set of basis states 
$\left\{|\Phi_i\rangle, \,i=1,\cdots, N\right\}$ which may be chosen as eigenstates of $H_0$. 
An eigenvector $|\Psi_1^{(N)}\rangle$ can be decomposed on this basis
\be
|\Psi_1^{(N)}\rangle = \sum_{i=1}^{N}  a_{1i}^{(N)}(g^{(N)})|\Phi_i\rangle
\label{eq2} \  
\ee    
 
Using the Feshbach projection method~\cite{fesh} ${\cal H}^{(N)}$  is decomposed into subspaces 
by means of the projection operators $P$ and $Q$, 
\be\nonumber
{\cal H}^{(N)} = P{\cal H}^{(N)} + Q{\cal H}^{(N)}
\ee

Here the subspace $ P{\cal H}^{(N)}$ is chosen to be of dimension 
$\mathrm{dim}\,P{\cal H}^{(N)}= N-1$ by elimination of one  basis state. The projected 
eigenvector $P|\Psi_1^{(N)}\rangle$ obeys the Schroedinger equation  
\be
H_{eff}(\lambda_1^{(N)})P |\Psi_1^{(N)}\rangle =  \lambda_1^{(N)}
P |\Psi_1^{(N)}\rangle
\label{eq3} \ . 
\ee
where $H_{eff}(\lambda_1^{(N)})$ operates in the subspace $P{\cal H}^{(N)}$. It is a nonlinear 
function of the eigenvalue $\lambda_1^{(N)}$ ~\cite{khri}, the eigenenergy being equal to the 
eigenvalue associated to $|\Psi_1^{(N)}\rangle$ in the initial space ${\cal H}^{(N)}$. In practice 
the coupling $g^{(N)}$ which characterizes the Hamiltonian $H^{(N)}$ in ${\cal H}^{(N)}$ is  
aimed to be changed in such a way that the eigenvalue in the  new space ${\cal H}^{(N-1)}$ is the 
same as the one in the complete space ${\cal H}^{(N)}$
\be
\lambda_1^{(N-1)} = \lambda_1^{(N)} 
\label{eq4} \ . 
\ee

Hence the reduction of the vector space from $N$ to $N-1$ results in a renormalization of 
the coupling constant from $g^{(N)}$ to $g^{(N-1)}$ preserving the physical eigenenergy 
$\lambda_1^{(N)}$, i.e. $\lambda_1^{(N)} = \lambda_1^{(N-1)} = \lambda_1$. The determination of 
$g^{(N-1)}$ by means of the constraint expressed by Eq.~(\ref{eq4}) is the central point of the 
procedure. If the dependence on $g$ is linear, i. e. $H = H_0 + g H_1$, it is obtained as a solution 
of an algebraic equation of the second degree in $g^{(N-1)}$, see Eq.(17) in Ref.~\cite{khri}. 

Starting from ${\cal H}^{(N-1)}$ the reduction procedure is iterated step by step in 
decreasing dimensions of the Hilbert space, $N \mapsto N-1 \mapsto N-2 \mapsto...$...At each step 
$g$ is renormalized to a new value. In the limit where the dimensions of ${\cal H}$ are very 
large one may go over to continuous space dimensions, $ (n,n-1) \mapsto (x,x-dx)$,  and the 
evolution of $g$ will be given by a flow equation, see Eq.(24) in Ref.~\cite{khri} and 
Ref.~\cite{jr}. The algorithm can be generalized to Hamiltonians depending on several coupling 
constants.

\subsubsection{Remarks.}

\begin{itemize}

\item The implementation of the reduction procedure asks for the knowledge of $\lambda_1$ and 
the corresponding eigenvector $|\Psi_1^{(k)}\rangle$ at any size $k$ of the vector space. The 
eigenvalue is in principle fixed as being the physical ground state energy of the system. We use 
the Lanczos algorithm which allows to determine $\lambda_1$ and $|\Psi_1^{(k)}\rangle$ 
~\cite{henk,lanc1,lanc2}. Consequently this algorithm has been implemented at each step of the 
reduction process~\cite{khri2}.\\
 
\item The process does not guarantee a rigorous stability of the eigenvalue $\lambda_1$. Indeed one 
notices that $|\Psi_1^{(k-1)}\rangle$ which is the eigenvector in the space ${\cal H}^{(k-1)}$ 
and the projected state $P|\Psi_1^{(k)}\rangle$ of $|\Psi_1^{(k)}\rangle$ into 
${\cal H}^{(k-1)}$may differ from each other. As a consequence it may not be possible to keep  
$\lambda_1^{(k-1)}$ rigorously equal to $\lambda_1^{(N)} = \lambda_1$. In practice the degree 
of accuracy depends on the relative size of the eliminated amplitudes $a_{1k}^{(k)}(g^{(k)})$.
We shall come back to this point when the algorithm will be implemented in numerical tests.

\item The reduction procedure needs a fixed ordering of the sequentially eliminated basis states. 
This ordering may be chosen by following different criteria. Here the states are arranged according 
to increasing energies $\epsilon_i = \langle\Phi_i^{(N)}|H|\Phi_i^{(N)}\rangle$ and eliminated 
starting from the one which corresponds to the highest energy at each step of the procedure.

\item The algorithm will be used in applications of the procedure on explicit models, here frustrated 
spin ladders, and tested in different symmetry schemes. As already mentioned we concentrate on aspects
related to systems close to phase transition points.

\end{itemize}
   
\subsection{Fixed points} 

The eigenvalues $\lambda_k{(g)}$ of $H(g) = H_0 + H_1(g)$ are analytic functions of $g$ which may 
show algebraic singularities~\cite{kat,hei,sch2} at so called exceptional points $g = g_e$. Exceptional
points are first order branch points in the complex $g$ - plane which appear when two (or more) 
eigenvalues get degenerate. This can happen if $g$ takes values such that $\epsilon_k =  \epsilon_l$ 
where $\epsilon_k  = \langle \Phi_k|H|\Phi_k\rangle$. In a finite Hilbert space the degeneracy appears
as an avoided crossing for real $g$. If an energy level $\epsilon_k$ belonging to the $P{\cal H}$ 
subspace defined above crosses an energy level $\epsilon_l $ lying in the complementary $Q{\cal H}$ 
subspace the perturbation development constructed from $H_{eff}(E)$ diverges~\cite{sch2}. 

Due to symmetry properties physical states can get degenerate in energy for real values of $g_e$. 
They correspond to first order phase transitions.

Exceptional points are defined as the solutions of ~\cite{hei,sch2}
\be 
f(\lambda(g_e))  = det[ H(g_e) - \lambda(g_e)I] = 0
\label{eq5}\ 
\ee
and 
\be
\frac{df(\lambda(g_e))}{d\lambda}|_{\lambda= \lambda(g_e)}= 0
\label{eq6}\ 
\ee

It is possible to show that exceptional points are fixed points of the coupling strength $g$ which
stay constant during the space reduction process. 

If $\left\{\lambda_i(g)\right\}$ are the set of eigenvalues the secular equation can be written 
as
\be
\prod_{i=1}^N {(\lambda - \lambda_i)} = 0
\label{eq7} \ .
\ee
Consider $\lambda = \lambda_p$ which satisfies Eq.~(\ref{eq5}). Eq.~(\ref{eq6}) can only be 
satisfied if there exists another eigenvalue $\lambda_q = \lambda_ p$, hence if a degeneracy 
appears in the spectrum. This is the case at an exceptional point. 

If the eigenvalue $\lambda_j^{(k)}, k= N, N-1,...$ which is either constant or constrained to take 
the fixed value $\lambda_j$ gets degenerate with some other eigenvalue 
$\lambda_i^{(k)}({g = g_e})$  in the space reduction process this eigenvalue must obey 
\be
\lambda_i^{(k)}({g_e}) = \lambda_i^{(l)}({g'_e})
\label{eq8} \ 
\ee
which is realised in any projected subspace of size $k$ and $l$ containing  states 
$|\Phi_j\rangle$ and $|\Phi_i\rangle$. Going over to the continuum limit for large values of $N$ 
as introduced above and considering the subspaces of dimension $x$ and $x + dx$  
 
\be
\lambda_i(g_e(x),x) = \langle\Psi_{i}(g_e(x),x)|H(g_e(x))|\Psi_{i}(g_e(x),x)\rangle \notag \\
\ee

verifies

\be
\frac{d\lambda_j}{dx} = 0 = \frac{d\lambda_i(g_e(x),x)}{dx}
\label{eq9} \ 
\ee 
 
Consequently
\be
\frac{\partial \lambda_i}{\partial x} + \frac{\partial \lambda_i}{\partial g_e} 
\frac{dg_e}{dx} = 0
\label{eq10} \ 
\ee

Since $\lambda_i(g_e(x),x)$ does not move with $x$ in the space dimension interval $(x,x+dx)$ the 
first term in Eq.~(\ref{eq10}) is equal to zero. Hence in general

\be
\frac{\partial \lambda_i}{\partial g_e}\not = 0 \,\,\, \quad {\mathrm{and}}\,\,\,\,\,
\frac{dg_e}{dx} = 0
\label{eq11} \ 
\ee  

The present result is general, it is valid for both continuous and discontinuous quantum phase 
transitions. Avoided crossings reflect the existence of continuous transitions in the limit of 
systems of infinite size. In the case of discontinuous (first order) transitions the energies of 
the physical states  $\{|\Psi_i\rangle\}$ show a real degeneracy point, i.e. their energies cross 
at real values of $g = g_{e}$ as already mentioned above.\\

In the following we shall consider Hamiltonians linear in $g$, $H(g) = H_0 + g H_1$. A  sufficient 
condition for possible level crossings is given by 

\ba
[H_0, H_1] = 0
\label{eq12} \
\ea
i.e. $H_0$ and $H_1$ can be simultaneously diagonalized~\cite{sach1,hei1}. In this case if 
$H_0$ is diagonal in the $\left\{|\Phi_i\rangle, \,i=1,\cdots, N\right\}$ basis of states 
it is also an eigenbasis for $H$ and

\ba
\frac{d\lambda_i(g_e(x),x)}{dx} = \frac{d\langle \Phi_{i}|H_0 + g_{e}(x)H_1|\Phi_ {i} \rangle}{dx}  
\label{eq13} \ 
\ea 

Since $H_0$ and $H_1$ do not depend on $x$, $d\lambda_i/dx = 0$  implies $dg_{e}/dx = 0$ for 
this specific form of the Hamiltonian which is consistent with the general result 
Eq.~(\ref{eq11}).\\

\subsection{Remarks.}

\begin{itemize}

\item  In the general case $H(g) = H_0 + H_1(g)$ $g(x) = g_{e}$ evolves through a flow equation 
which is more complicated than Eq. (24) in ~\cite{khri}. We shall restrict our numerical investigations
to Hamiltonians which show a linear dependence on $g$.

\item The energy level crossings can occur for any two energy eigenstates of the spectrum. We 
shall consider level crossings or avoided crossings between the ground state and an excited state 
as well as the case of crossings or avoided crossings of excited states. 

\end{itemize}

The present formal considerations are now used in numerical applications. We analyze the 
behaviour of the reduction algorithm in the neighbourhood of fixed points and show to what 
extent it allows to detect fixed points in an application to two-legged frustrated spin
ladders for different choices of the basis of states.

\section{Application to quantum transitions in frustrated two-leg quantum spin ladders.}

\subsection{The model}

\begin{figure}
\epsfig{file=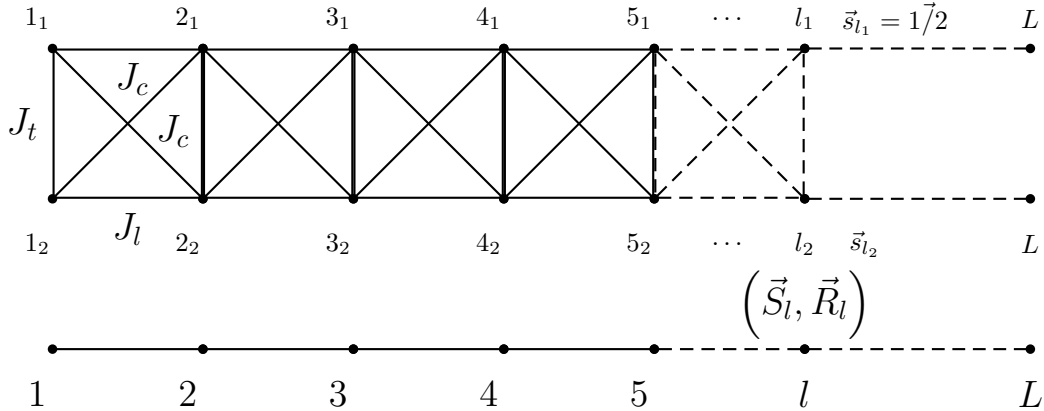}
\caption{Top: the original spin ladder. The coupling strengths are indicated as given in the text. 
Bottom: The ladder in the SO(4) representation. See the text.}
\label{fig0}
\end{figure}

\subsubsection{SU(2)-symmetry framework.}

Consider spin-$1/2$ ladders~\cite{lin1,lin2} shown in Fig. 1 and described by Hamiltonians of the 
following type 
\ba
H^{(s,s)} &=& J_t\sum_{i=1}^{L} s_{i_1}s_{i_2} + J_l\sum_{<ij>}s_{i_1}s_{j_1} +
J_l\sum_{<ij>}s_{i_2}s_{j_2} \notag \\ 
&  &   + J_{1c}\sum_{(ij)}s_{i_1}s_{j_2} + J_{2c}\sum_{(ij)}s_{i_2}s_{j_1} 
\label{eq14}
\ea
 
Working in a representation with fixed total magnetic projection $M_{tot}$ the basis of states is 
spanned by the vectors $\left\{|\Phi_k\rangle, \,k=1,\cdots, N\right\}$ where 

\be\nonumber
|\Phi_k\rangle = |1/2 ~~m_1,...,1/2 ~~m_i,...,1/2 ~~m_{2L}, \sum_{i=1}^{2L} m_i= M_{tot}\rangle  
\ee
and $\{m_i= +1/2, -1/2\}$.

The indices $1$ or $2$ label the spin $1/2$ vector operators $s_{i_k}$ acting on the sites $i$ 
on both ends of a rung, in the second and third term $i$ and $j$ label nearest neighbours, here 
$j = i+ 1$ along the legs of the ladder. The fourth and fifth term correspond to diagonal 
interactions between nearest sites located on different legs. $2L$ is the number of sites on 
a ladder. Here we fix  $J_{1c} = J_{2c} = J_c$. The coupling strengths $J_t, J_l, J_{c}$ are 
positive. In the sequel we restrict our analysis to the case where $M_{tot} = 0$.\\

In the present applications the renormalization is restricted to a unique coupling strength, 
see Eq.~(\ref{eq1}). It is implemented here by putting $H_0 = 0$ and $H^{(N)} = g^{(N)} H_1$ where 
$g^{(N)} = J_t$ and
\ba
H_1 &=& \sum_{i=1}^{L} s_{i_1}s_{i_2} + \gamma_{tl} \sum_{<ij>}(s_{i_1}s_{j_1} +
s_{i_2}s_{j_2})\notag \\   
& & + \gamma_{c}\sum_{<ij>}(s_{i_1}s_{j_2}+s_{i_2}s_{j_1}) 
\label{eq15} \  
\ea
where $\gamma_{tl} = J_{l}/J_{t}$, $\gamma_{c} = J_{c}/J_{t}$. These quantities are 
kept fixed and $g^{(N)} = J_t$  will be subject to renormalization in the reduction process.\\

\subsubsection{SO(4)-symmetry framework.}

The basis of states may be written in an $SO(4)$-symmetry scheme. By means of a spin rotation 
~\cite{kika,kimo}

\be
 s_{i_1} = \frac{1}{2} (S_i + R_i)
\label{eq16} \ .
\ee
\be
 s_{i_2} = \frac{1}{2} (S_i - R_i)
\label{eq17} \ .
\ee
the Hamiltonian Eq.(15) can be expressed in the form\\
 
\ba
H^{(S,R)} &=& \frac{J_t}{4}\sum_{i=1}^{L} (S_{i}^{2} -  R_{i}^{2}) + J_1\sum_{<ij>}S_i S_j 
\\ \nn
&  & + J_2\sum_{<ij>}R_i R_j
\label{eq18}
\ea
which reduces the ladder formally to a chain, see bottom of Fig. 1. Here
$J_{1}=(J_{l}+J_{c})/2$, $J_{2}=(J_{l}-J_{c})/2$, $J_{1c}=J_{2c}=J_{c}$. The components 
$S_{i}^{(+)}, S_{i}^{(-)}, S_{i}^{(z)}$ and $R_{{i}}^{(+)}, R_{{i}}^{(-)},R_{{i}}^{(z)} $ of 
the vector operators $S_{i}$ and  $R_{i}$ are the $SO(4)$ group generators and $<ij>$ denotes 
nearest neighbour rung indices
 
\ba\nonumber
S_{i}^{(+)}  = \sqrt{2}(X^{(11)(10)}_{i} + X^{(10)(1-1)}_{i}) = S_{i}^{(-)*}
\ea
 
\ba\nonumber
S_{i}^{(z)}  =  X^{(11)(11)}_{i} - X^{(1-1)(1-1)}_{i}
\ea

\ba\nonumber
R_{i}^{(+)}  =  \sqrt{2}(X^{(11)(00)}_{i} - X^{(00)(1-1)}_{i}) = R_{i}^{(-)*}
\ea
 
\ba\nonumber
R_{i}^{(z)} =  - (X^{(10)(00)}_{i} + X^{(00)(10)}_{i})
\ea 

where 
\ba\nonumber
X^{(S_i M_i)(S'_i  M'_i)}_{i} = |S_i M_i \rangle  \langle S'_i  M'_i|\nonumber 
\ea

with states defined as  
\ba\nonumber
|S_i M_i \rangle = \sum_{m_1,m_2} \langle 1/2 ~~m_1 ~~1/2 ~~m_2|S_i M_i  \rangle
|1/2 ~~m_1\rangle_{i}  |1/2 ~~m_2\rangle_{i}
\ea 

along a rung are coupled to $S_i = 0$ or $S_i = 1$. Spectra are constructed in this representation 
as well as in the $SU(2)$ representation. 

In a representation with fixed total magnetic projection $M_{tot} = 0$ the basis of states is then 
spanned by the vectors 
\be\nonumber
|\Phi_k\rangle = |S_1 M_1,...,S_i M_i,...,S_{L} M_{L}, \sum_{i=1}^{L} M_i= M_{tot} = 0\rangle  
\ee
and $\{M_i= 0, +1, -1\}$\\

In the sequel  we analyse the behaviour of the system in both symmetry frames.

\subsection{Fixed points in the SU(2)-symmetry framework.} 
 
\subsubsection{First order phase transitions of the two-legged spin ladder.}

At first order transitions which happen at level crossings for real $g$ the amplitudes
of the wavefunctions are expected to acquire weights of the same order of magnitude over a
large number of basis states. The analyses performed in ~\cite{khri2} show that the use
of the algorithm at fixed points should work as a stringent test of the method. 

The first application concerns a system described by the Hamiltonian $H^{(s,s)}$ 
given by Eq.~(\ref{eq14}) using an $SU(2)$-symmetry basis of states given below 
Eq.~(\ref{eq14}). A crossing between a rung dimer phase and a Haldane phase appears 
for $J_l = J_{c}$ when $J_t/J_l \simeq 1.401$ in the case of an asymptotically 
large system~\cite{lin1,lin2,gel,wko,bos,wan}. The ratio depends on the size of the system. 
The existence of a first order transition is analysed below, the coupling constant 
$g = J_{t}$ is expected to stay constant at the level crossing point.

\subsubsection{Application of the reduction algorithm at fixed points.}

We consider ladders with $L=6$. Several crossings between energy levels can be observed in 
Fig. 2(a) which shows the evolution of the energies of the four lowest states in the 
$M_{tot} = 0$ subspace as a function of $g = J_{t}$. The crossing between the ground state 
energy $e_1$ and the energy of the first excited state $e_2$ corresponds to $J_t/J_l \simeq 1.23$. 

The first test corresponds to $J_{t} \ge 6$. Then the three lowest excited states corresponding 
to $e_2, e_3, e_4$ get rigorously degenerate which generates a continuous transition line~\cite{wko}.
Fixing $J_{t} = 10$  Fig. 2(b) shows the evolution of the four lowest states as a function of the 
size $N$ of the Hilbert space following the algorithm described in Refs.~\cite{khri,khri2}. 
The initial space dimension is $N = 924$. The stability of the spectrum is remarkable down to 
$N \leq 100$. This stability reflects in the constancy of $J_{t}$ over the same dimensional range, 
see Fig. 2(c). For lower values of $N$ deviations appear. They are due to the approximations inherent 
to the projection method as noted in the second remark of subsection 2.1.3 and Refs.~\cite{khri,khri2}. 
The limit of constancy of $J_t$ indicates the minimum dimension of Hilbert space in which diagonalization 
will lead to the reproduction of the low energy part of the spectrum, i. e. the ground and first 
excited states. The behaviour of the spectrum  shown here is observed for any value of $J_{t} \ge 6$, 
i. e. all along the degeneracy lines.

Fig. 2(d) gives the evolution of the entropy $s$ per site defined as ~\cite{zel}  

\ba
s = - \frac{1}{2L} \sum_{i=1}^{N}{P_i}ln{P_i} & with & P_i = |\langle\Phi_i^{(N)}|\Psi_1^{(N)}
\rangle|^{2} = |a_{1i}^{(N)}|^{2}
\label{19}
\ea
at the fixed point $J_{t} = 5$ which corresponds in Fig. 2(a) to the crossing of the ground state 
$e_1$ with the first excited state $e_2$. Here $\{a_{1i}^{(N)}\}$ are the amplitudes of the 
components of the ground state wavefunction developed on the basis of states $\{|\Phi_i>\}$ which 
span the space of dimension $N$. The step discontinuity signals the transition characterized by a 
strong change in the structure of the lowest state. The characteristic singularity observed at this 
value of $J_t$ is conserved as long as the ground state keeps stable during the space reduction process.

At the exact location of the fixed point the instability of the spectrum is sizable and the coupling 
constant $J_{t}$ at this crossing point stays constant over a smaller interval of values of $N$.
A closer inspection shows that this instability might be related to a numerical difficulty in the 
renormalization of $J_{t}$. Indeed the coefficients of the algebraic second order equation which fixes 
it ~\cite{khri} get accidentally vanishingly small at this place and consequently lead to strongly
unprecise values of the roots of the equation. This corresponds to a pathological situation 
which may not be significative in the general case. Indeed, in the close neighbourhood of the fixed point, 
$J_t$ stays stable over a much larger interval of values of $N$ when $J_c = 3.8 \not= J_l$ 
(see Figs. 2 (e) and 2(f)).

\subsection{Fixed points in the SO(4)-symmetry framework - First order transitions}

The reduction algorithm is next applied to the same system as above but described by the Hamiltonian 
$H^{(S,R)}$ given by Eq. (19) with a basis of states $\{|\Phi_i>\}$ written in the $SO(4)$ 
symmetry framework introduced in subsection 3.1.2. The spectrum given in 
Fig. 3(a) in the $M_{tot} = 0$ subspace is the same as in the case of the $SU(2)$ symmetry 
framework as it should be. However the behaviour of the numerically generated spectrum at different 
transition points is quite different. 

At the crossing point between the ground state and the first excited state in the $M_{tot} = 0$ 
subspace which occurs at $J_{t} =5$ (see Fig. 3(a)) the ground state energy $e_1$ and $J_{t}$ remain 
stable all along the space dimension reduction procedure as seen in Figs. 3(b) and 3(c). This 
remarkable stability can be explained by the fact that the ground state wavefunction is strongly 
dominated by a small number of states in the $SO(4)$ basis. It is however lost for the first 
excited state with energy $e_2$ which moves abruptly and stays then again constant generating 
successive plateaus over more or less large intervals in $N$, Fig. 3(b). This shows that 
$d\lambda_i/dx = 0$, $(i = 2,3,4)$ is indeed preserved by steps, but not necessarily $\lambda_i$ 
which jumps by steps over finite intervals of space dimensions. The jumps in $\{\lambda_i\}$ may 
be related to the elimination of non negligible components of the wavefunction during the reduction 
process. 

Fig. 3(d) shows the behaviour of $s$ the ground state entropy per site which behaves like in the 
$SU(2)$ scheme but is quantitatively smaller. This is due to the fact that the wavefunction amplitudes
are less equally distributed here than in the $SU(2)$-scheme as mentioned above. Figs. 3(e) and 3(f) 
show the behaviour of the spectrum and coupling constant $J_t$ in the close neighbourhood of the 
transition point. One observes that the evolution of the energies is smoother than at the transition 
point itself and the coupling constant increases slightly with decreasing $N$. Some curves in the 
figures are drawn with a finite width in order to facilitate the observation of the stepwise 
evolution of the corresponding quantities. Degeneracy of the states and the consequent constancy 
for $J_{t} \ge 6$ is observed along $e_2 = e_3 = e_4$ which corresponds to a transition line as 
seen in Figs. 4(a) and 4(b). The constancy of these quantities is preserved over the whole range 
of space dimensions $N$, except for $\{e_i\}$'s at small $N$. But $\{e_i\}$'s stay more and more 
constant up to the smallest values of $N$ with increasing $J_t$, see Figs. 4(c) and 4(d). This can 
be explained by the fact that the wavefunctions gets more and more dominated by a small number of 
basis states with increasing $J_t$. Evidently the robustness of the spectrum is stronger 
in the present symmetry scheme than in the case of $SU(2)$.\\ 
   
\subsection{Application of the reduction algorithm at a continuous transition: avoided crossings.}

As already mentioned continuous transitions reduce to avoided crossings in finite systems. 
States get degenerate at complex values of this parameter. Genuine transitions with real 
parameters cannot be explicitly seen in numerically determined spectra of finite systems. 
Avoided crossings are not easy to locate. They occur at complex exceptional points and the 
present numerical procedure is aimed to follow the evolution of real running coupling constants. 

Fig. 5(a) shows the spectrum of the ladder for specific values of the coupling constants. One
observes several possible avoided crossings which are rather close to each other in the 
interval $4<J_t<9$. The typical behaviour of the spectrum is shown in Fig. 5(b) for 
$J_t=6.6891$. Fig. 5(c) gives a quantitative estimate of the energy fluctuations of excited 
states. Here

\ba\nonumber
p(i) = |\frac{(e_i^{(N)}-e_i^{(N-k)})}{e_i^{(N)}}| \times 100 & with& i=1,\ldots,4
\ea  

The spectrum and $J_t$ are relatively stable down to $N \simeq 200$. Stability is lost below this value.
The same is true at other avoided crossing points. These points are difficult to locate, the coupling
$J_t$ is complex there and our procedure does not fix their imaginary part. It may be that clearer 
signals can be observed for larger systems since then the gap at crossings gets smaller (in principle 
tends to zero in case of a continuous transition) and hence leads to a reduced the imaginary part of 
the coupling constant.

Remarks:

\begin{itemize}
 
\item In the present calculations the system has open boundary conditions. For odd $L$ the entropy
shows the same characteristic discontinuity as observed for first order transitions in Figs. 2(d) and 
3(d) at the transition point. For $L$ even the discontinuity goes over in a finite peak which recalls 
a continuous transition in a finite size system. This shows that a naive interpretation of observables 
in finite systems can lead to erroneous interpretations of the order of a transition.

\item Expected crossings may not necessarily signal continuous transitions in the limit of infinite 
systems. In this limit the order may effectively be different. This phenomenon has also been observed 
in classical systems see f.i.~\cite{ple} and references therein.

\end{itemize}

\section{Summary, conclusions and outlook.}

We used an algorithm which aims to reduce the size of the Hilbert space of states describing 
strongly interacting systems. The reduction induces a renormalization of coupling constants which 
enter the Hamiltonians of the systems, here frustrated two-leg quantum spin ladders. The robustness 
of the algorithm has already been tested in former work~\cite{khri2}.

We applied the algorithm at the location and in the neighbourhood of first order transition 
points and lines, and in the vicinity of avoided crossings which correspond to second order 
transitions in infinite systems. The analysis has been pursued in two different symmetry schemes. 
As it may be expected the behaviour of the spectrum and the renormalized coupling parameter depend 
on the symmetry framework. Indeed, the description of the system depends crucially on the details 
of the wavefunctions of the different energy states and their structure is different in different 
representations.    

In the case of first order transitions we showed that the renormalized coupling constant $J_t$ may
indeed be numerically stable over a large set of dimensions of Hilbert space and signal the presence 
of a transition as predicted by the theory considerations developed in section 2.2. Strong instabilities
may appear due to accidental numerical pathologies as mentioned in section 3.2.2.

The stability is larger in the case of an $SO(4)$-symmetry representation where the number of 
large components is reduced. In the present parameter space the low-lying eigenstates are strongly 
dominated by a small set of $SO(4)$ basis states. This may also explain the stronger stability of 
degenerate states at the crossing points and along transition lines.  

Avoided crossings have also been investigated. The presence of these points is difficult to locate. 
It may be due to the fact that the level crossing point occurs for a complex coupling constant 
which is not detected in the present algorithm. The precise understanding of these points related 
to numerics requires further work which lies outside the scope of the present investigations.

It may be mentioned here that there exists other methods which are aimed to detect crossing and 
avoided crossing points. One of them relies on discontinuities in the entanglement properties of  
wavefunctions ~\cite{wu,oli} which have to be known at crossing (first order transitions) or 
avoided crossings  (continuous transitions). A second approach relies on an algebraic method 
~\cite{bha} which works very nicely in the case of small systems. It is not clear however whether 
its use can be easily applied to very large Hilbert spaces such as those which correspond to
realistic quantum spin systems. 

In summary, the present investigations show that first order transition points in quantum spin 
systems may be correlated with strong fluctuations in the energies of low-lying excited states. The
presence of these points are signalized by the constancy of the strength of the couplings which 
enter the Hamiltonian along the dimensional reduction procedure of Hilbert space. This is predicted by 
theoretical considerations and verified, at least up to the point at which numerical stability gets 
lost which may happen at different stages of the reduction procedure depending on the structure  
of the ground state wavefunction.
 
One of us (T.K.) would like to thank Drs. T. Vekua and R. Santachiara for critical remarks and 
helpful advice.

\newpage

\begin{figure}[b]
\includegraphics{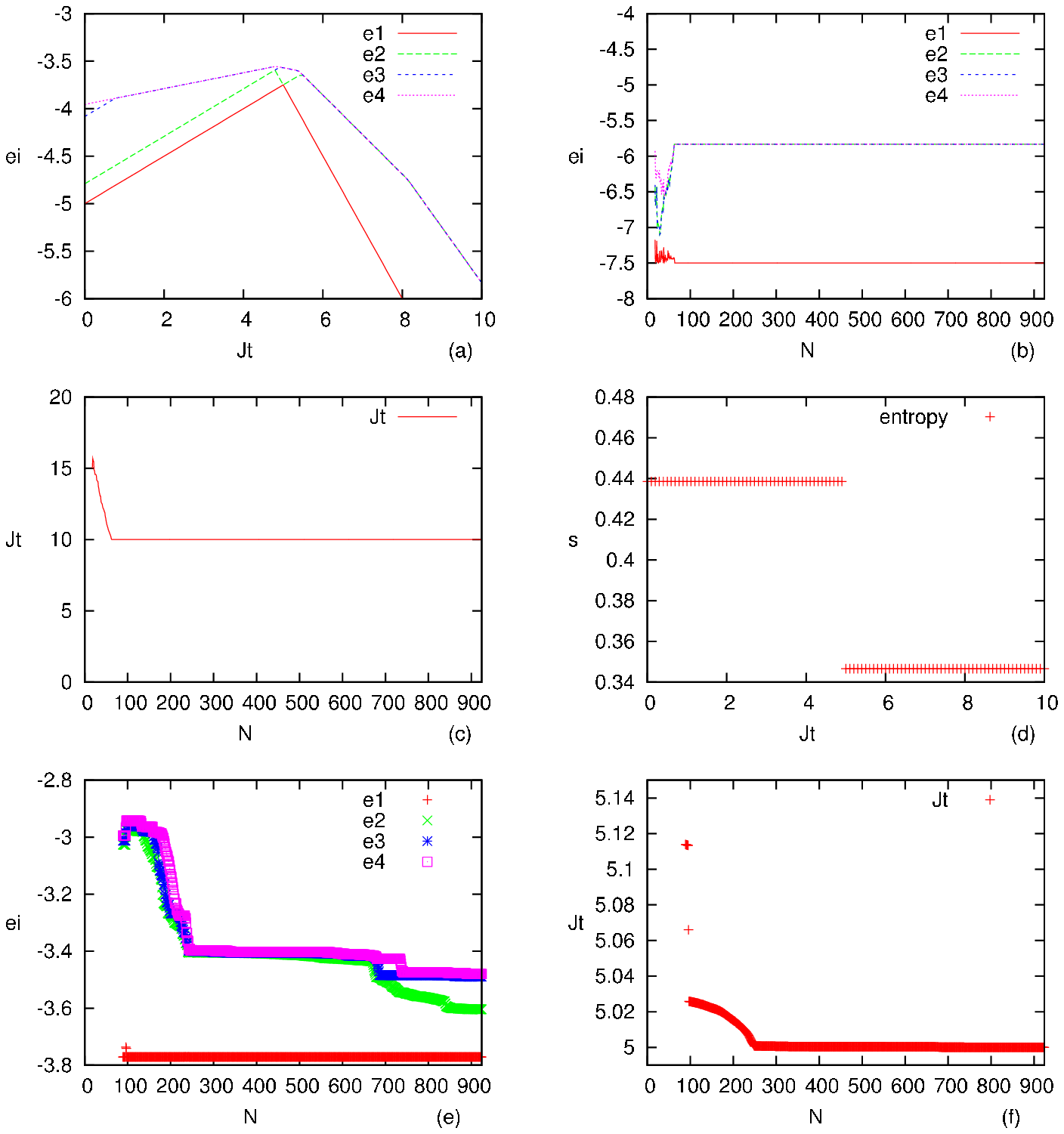}
\caption{$SU(2)$-symmetry  scheme. The $\{e_i, i= 1,2,3,4\}$ are the energies per site of the 
ground and lowest excited states. $N$ is the size of the Hilbert space, $s$ the entropy of the 
ground state per site. The number of sites is $L=6$ along a leg, $J_l=J_{c} \simeq 4.07$. 
(b) and (c) correspond to $J_t = 10$. Figs.(e) and (f) correspond to $J_c = 3.8 \not= J_l$.  
Broadened lines are drawn for the sake of readableness. See discussion in the text.}
\label{fig1}
\end{figure}

\begin{figure}[b]
\includegraphics{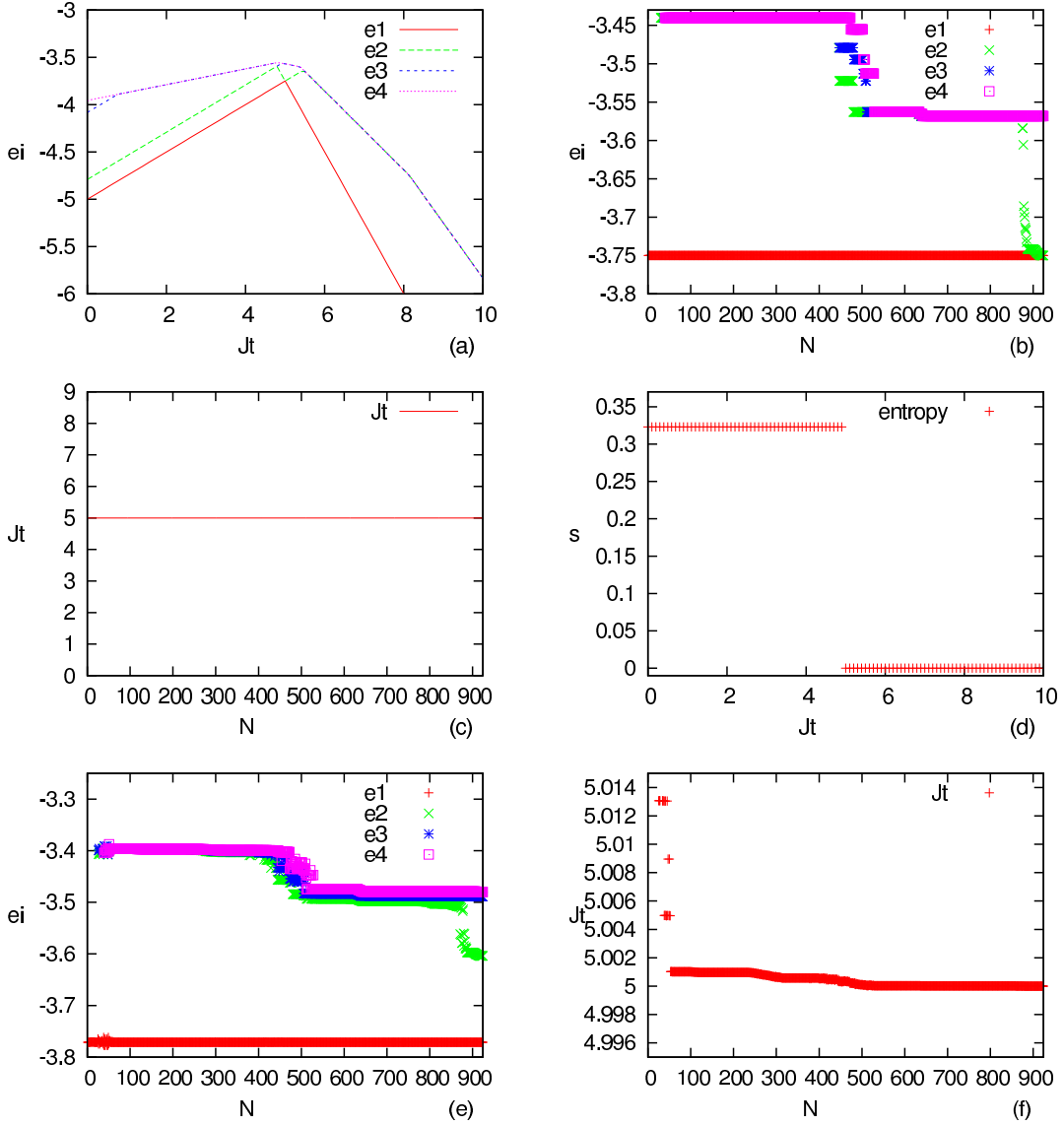}
\caption{$SO(4)$-symmetry  scheme. The $\{e_i, i= 1,2,3,4\}$ are the energies per site of the 
ground and lowest excited states. $N$ is the size of the Hilbert space, $s$ the entropy of the 
ground state per site. The number of sites is $L=6$ along the chain. 
(a) - (d): $J_l=J_{c}$. In (b) and (c) $J_t = 5$
(e) - (f): $J_t = 5$,  $J_{l} \neq J_{c}$, $J_{c} = 3.8$.  
In both cases $J_{t}/J_{l} \simeq 1.23$. Broadened lines are drawn for the sake of readableness.
See discussion in the text.}
\label{fig2}
\end{figure}

\begin{figure}[b]
\includegraphics{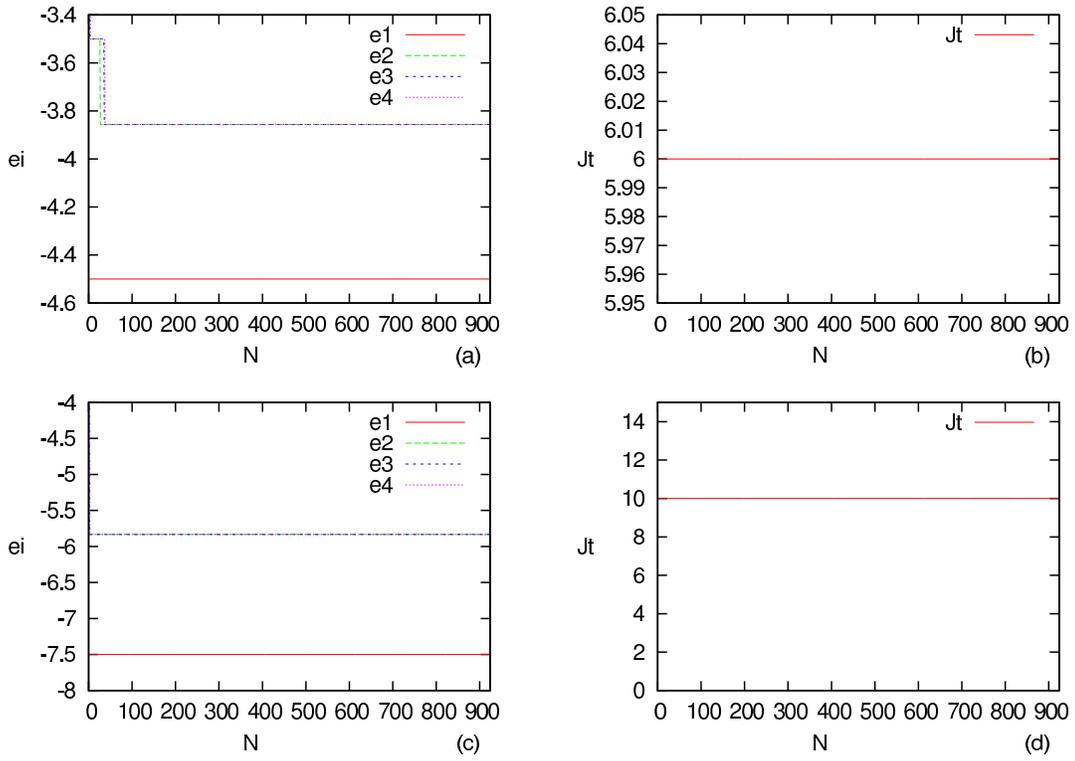}
\caption{$SO(4)$-symmetry  scheme. The $\{e_i, i= 1,2,3,4\}$ are the energies of the ground 
and lowest excited states per site. The number of sites is $L=6$ along the chain, $J_l=J_{c}
\simeq 4.07$. (a) and (b) correspond to $J_{t} = 6$, (c) and (d) correspond to $J_{t} = 10$. 
See discussion
in the text.} 
\label{fig7}
\end{figure}

\begin{figure}[b]
\includegraphics{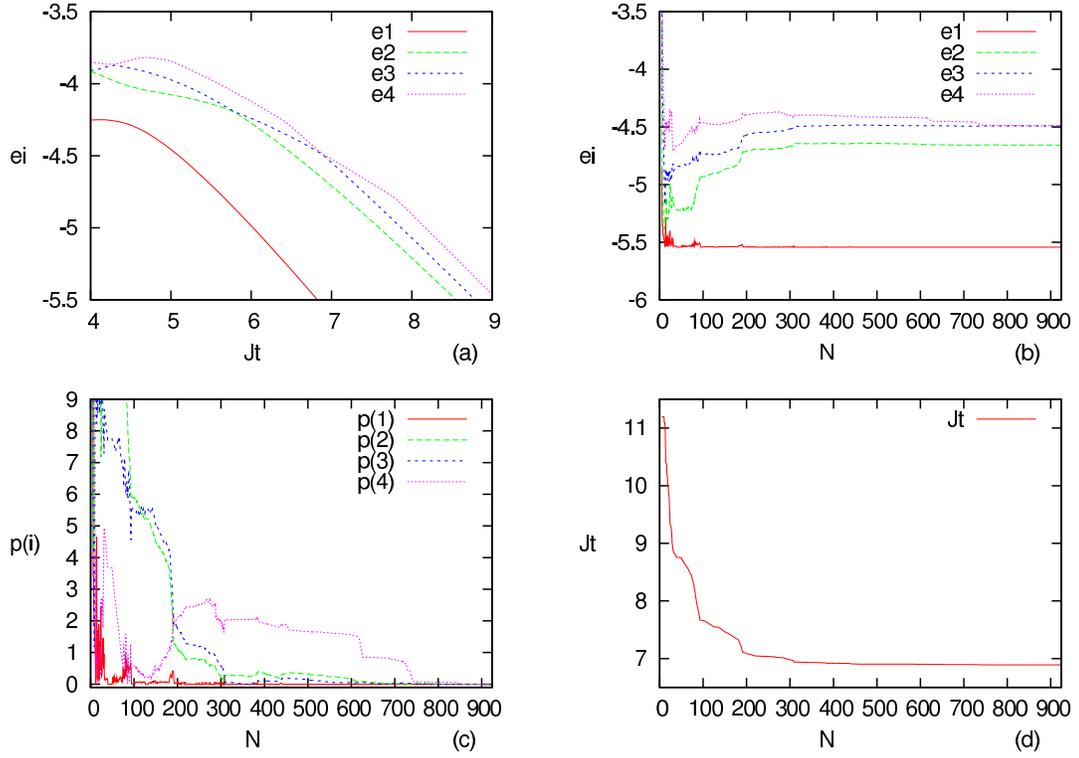}
\caption{$SU(2)$-symmetry  scheme. The $\{e_i, i= 1,2,3,4\}$ are the energies per site of the ground 
and excited states. $N$ is the size of the Hilbert space.
 The number of sites is $L=6$ sites, $J_t=6.891$, $J_l=5,J_{c}=3$. See discussion in the text} 
\label{fig6}
\end{figure}

\end{document}